**Magnetic Domain-Wall Tilting due to Domain-Wall Speed Asymmetry**

Dae-Yun Kim[1], Min-Ho Park[1], Yong-Keun Park[1], Joo-Sung Kim[1,2], Yoon-Seok Nam[1], Hyun-Seok Hwang[1], Hyeok-Cheol Choi[1], Duck-Ho Kim[3], Soong-Geun Je[4], Byoung-Chul Min[2], and Sug-Bong Choe[1,†]

[1]Department of Physics and Institute of Applied Physics, Seoul National University, Seoul, 08826, Republic of Korea.

[2]Center for Spintronics, Korea Institute of Science and Technology, Seoul, 02792, Republic of Korea.

[3]Institute for Chemical Research, Kyoto University, Kyoto, 611-0011, Japan

[4]CNRS, SPINTECH, F-38042 Grenoble, France

[†]Correspondence to: sugbong@snu.ac.kr

**Chiral magnetic materials provide a number of challenging issues such as the highly-efficient domain-wall (DW) and skyrmion motions driven by electric current [1-4], as of the operation principles of emerging spintronic devices [5-7]. The DWs in the chiral materials exhibit asymmetric DW speed variation under application of in-plane magnetic field [8, 9]. Here, we show that such DW speed asymmetry causes the DW tilting during the motion along wire structure. It has been known that the DW tilting can be induced by the direct Zeeman interaction of the DW magnetization under application of in-plane magnetic field [10, 11]. However, our experimental observations manifests that there exists another dominant process with the DW speed asymmetry caused by either the Dzyaloshinskii–Moriya interaction (DMI) [9] or the chirality-dependent DW speed variation [12-17]. A theoretical model based on the DW geometry reveals that the DW**

**tilting is initiated by the DW pinning at wire edges and then, the direction of the DW tilting is determined by the DW speed asymmetry, as confirmed by a numerical simulation. The present observation reveals the decisive role of the DW pinning with the DW speed asymmetry, which determines the DW geometry and consequently, the dynamics.**

The DMI—an antisymmetric exchange interaction [18, 19]—has been actively studied nowadays due to its important role in spintronics such as formation of the magnetic skyrmions [3, 4, 20] or stabilization of the chiral domain-walls (DWs) [8, 21]. To quantify the DMI, several techniques have been developed based on the measurements of asymmetric DW speed [9, 22], the current-induced DW motion [23, 24], asymmetric hysteresis of magnetic patterns [25], and Brillouin light scattering [26]. Fairly recently, a theoretical study proposed another DMI measurement scheme based on the DW tilting in magnetic wire structure under application of in-plane magnetic field [10, 27]. It is obvious that, due to the Zeeman interaction between the in-plane magnetic field and the magnetization inside the DW, the DW has to be tilted to an equilibrium angle of minimum energy configuration. The equilibrium angle is determined by the counterbalance between the DMI and Zeeman energies. The present work was originally motivated to verify this prediction experimentally, but interestingly, we found the existence of another governing mechanism that overwhelms the prediction based on the energy minimization.

**Results**

**Observation of DW tilting**

Figure 1(a) plots the DW speed $v_{\text{DW}}$ with respect to the in-plane magnetic field $H_x$ for

Sample I, where $H_\perp$ is applied normal to the DW (for sample details, see methods). The plot clearly shows that $v_{DW}$ has a minimum at $H_0$ (blue vertical line) and exhibits a symmetric variation around $H_0$. According to Refs. [8, 9], the value of $H_0$ is a direct measure of the DMI-induced effective magnetic field $H_{DMI}$, since the minimum $v_{DW}$ appears at the Bloch-type DW configuration under zero total magnetic field i.e. $H_{DMI} + H_0 = 0$. Therefore, from the present measurement, $H_{DMI}$ is quantified as $-120\pm5$ mT in the direction normal to the DW. Due to the large negative $H_{DMI}$, the magnetization $M_{DW}$ inside the DW is aligned parallel to $H_{DMI}$, forming the Néel-type DW configuration as depicted in Fig. 1(b). At this instant, if one applies an additional in-plane magnetic field $H_y$ in the direction transverse to the wire, the Zeeman interaction will rotate $M_{DW}$, followed by tilting of the overall DW to keep the Néel-type DW configuration [10]. For the present sample with a negative $H_{DMI}$, it is expected that the DW has to rotate clockwise under application of $H_y$ ($> 0$) as shown by Fig. 1(c).

However, the experimental observation is opposite to the above prediction. Figure 1(d) shows the DW images after each successive application of out-of-plane magnetic field $H_z$ pulses. The first three images were taken without application of $H_y$ and the last three images were taken under application of $H_y$. The figure clearly shows the DW tilting between these two sorts of images, conforming that the tilting is truly caused by application of $H_y$ ($> 0$). However, it is surprising to note that the direction of the DW tilting is counterclockwise, which is opposite to the prediction shown by Fig. 1(c). The present experimental observation, therefore, indicates that there should exist another hidden governing factor of the DW tilting.

**DMI-induced DW tilting with edge pinning**

It is also interesting to note that the DWs with $H_y = 0$ show a typical shape of circular arcs. Such typical DW shape indicates the existence of pinning at the edges of wire [28, 29]. The subsequent effect from such arc formation can be explained as follows. Figure 2(a) depicts the typical DW shape of a circular arc. Due to the shape of arc, if one applies $H_y$ (blue arrows), the component $H_\perp$ (black arrows) normal to the DW varies on position. Then, due to the $v_{DW}$ dependence on $H_\perp$ as shown by Fig. 1(a), $v_{DW}$ also varies accordingly. For this sample, the upper and lower parts of the DW have positive and negative $H_\perp$, respectively, as shown by Fig. 2(a). Therefore, the upper part moves slower than the lower part and consequently, the overall DW shape rotates counterclockwise. One can therefore conclude that the pinning effect can trigger the DW tilting, of which the direction accords to the experimental observation.

For better insight, we consider the role of $H_{DMI}$ in the above process. Due to the shape of the arc, the direction of $H_{DMI}$ (red arrows) varies on position to keep the direction normal to the DW. If one applies $H_y$ here, due to the different angles between $H_{DMI}$ and $H_y$, $M_{DW}$ should rotate differently, resulting in non-uniform distribution of the DW energy density $\sigma_{DW}$. According to Ref. [8, 9], $\sigma_{DW}$ is given as

$$\sigma_{DW} = \sigma_0 + 2\lambda K_D \cos^2 \psi - \pi \lambda M_S [(H_\perp + H_{DMI}) \cos \psi + H_\parallel \sin \psi], \quad (1)$$

with the Bloch-type DW energy density $\sigma_0$, the DW width $\lambda$, the DW anisotropy energy density $K_D$, and the saturation magnetization $M_S$, where $H_\parallel$ is the component of $H_y$ parallel to the DW and $\psi$ is the angle of $M_{DW}$ from the direction normal to the DW. For general case, the equilibrium $\psi$ is determined by the energy minimization condition for a given $H_y$.

Though there is no explicit analytic solution of Eq. (1) with the equilibrium $\psi$, one can intuitively estimate $\sigma_{DW}$ for the case of a large negative $H_{DMI}$ as of Sample I, since $\cos \psi$ is almost set to $-1$ and does not change much under the experimental range of $H_y$.

For this case, one can readily obtain a simplified relation

$$\sigma_{\text{DW}} \approx \sigma_0 + \pi\lambda M_S(H_\perp + H_{\text{DMI}}). \tag{2}$$

Then, according to Ref. [9], $v_{\text{DW}}$ depends on $\sigma_{\text{DW}}$ via the DW creep criticality

$$v_{\text{DW}} = v_0 \exp\left[-\alpha(\sigma_{\text{DW}}/\sigma_0)^{1/4} H_z^{-1/4}\right], \tag{3}$$

where $v_0$ is the characteristic speed and $\alpha$ is the scaling constant [30]. These relations can explain the $v_{\text{DW}}$ dependence on $H_\perp$: a larger $H_\perp$ has a larger $\sigma_{\text{DW}}$ and consequently, a smaller $v_{\text{DW}}$ in accordance to Fig. 1(a). One can therefore reach the same conclusion that the upper part of the DW with positive $H_\perp$ moves slower than the lower part with negative $H_\perp$ and consequently, the overall DW exhibit counterclockwise tilting.

To confirm the present prediction, we carried out a numerical simulation based on Eqs. (1) and (3) by including consideration of the local pinning force $f_{\text{pin}}$. In the simulation, $f_{\text{pin}}$ is given as

$$f_{\text{pin}} = \begin{cases} f_{\text{film}} + \beta(y)f_{\text{edge}} & \text{near edges} \\ f_{\text{film}} & \text{otherwise} \end{cases}, \tag{4}$$

where $\beta(y) < 1$, to mimic the edge pinning qualitatively. Since $\alpha$ is proportional to $f_{\text{pin}}$ [30-32], one can replace $\alpha$ by $\alpha_0 f_{\text{pin}}$, where $\alpha_0$ is a constant irrespective of $f_{\text{pin}}$. Then, $v_{\text{DW}}$ depends now on $f_{\text{pin}}$. Figure 2(b) shows the simulation results for the case that $f_{\text{pin}}$ at the edge is given by 12 % larger than $f_{\text{pin}}$ at the center of the wire (for details, see Supplementary I). It is clear from the figure that the counterclockwise DW tilting is well reproduced in accordance to the experimental observations.

It is worthwhile to note that, for the case of a large positive $H_{\text{DMI}}$, $\cos\psi$ is set to $+1$ and thus, the previous $v_{\text{DW}}$ dependence on $H_\perp$ becomes reversed, resulting in clockwise

DW tilting. Therefore, one can conclude that the DW tilting itself is triggered by the edge DW pinning, but the direction of the DW tilting is determined by the sign of $H_{\text{DMI}}$.

It is also experimentally confirmed that, for the case that $H_{\text{DMI}} = 0$, the DW tilting does not take place, as expected. Figure 3(a) plots $v_{\text{DW}}$ with respect to $H_\perp$ for the Sample II (for sample details, see methods). The result shows that $H_0$ for minimum $v_{\text{DW}}$ is almost zero, revealing that $H_{\text{DMI}} \approx 0$ for this sample. Due to the symmetric $v_{\text{DW}}$ variation around $H_\perp = 0$, the lower and lower parts of the DW moves with the same speed irrespective of the sign of $H_\perp$. Therefore, the DW keeps basically the same shape of the symmetric circular arcs even after application of $H_y$ as seen in Fig. 3(b).

**DW tilting by chirality-induced $v_{\text{DW}}$ asymmetry**

In addition to the DMI-induced $v_{\text{DW}}$ variation, many groups have recently uncovered the existence of the chirality-induced $v_{\text{DW}}$ variation [12-17]. Here, we explore the effects of such chirality-induced $v_{\text{DW}}$ variation on the DW tilting mechanism. For this purpose, the DW tilting is observed from a sample that has nearly zero $H_{\text{DMI}}$, but exhibits sizeable chirality-induced $v_{\text{DW}}$ variation.

Figure 4(a) plots the spin-torque efficiency $\varepsilon_{\text{ST}}$ with respect to $H_\perp$ for Sample III (for sample and measurement details, see methods and Supplementary II). Since the plot shows the typical spin-orbit-torque-driven behavior [23], the intercept (red vertical arrow) to the abscissa axis directly quantifies $H_{\text{DMI}}$ [23, 33]. The plot thus indicates that $H_{\text{DMI}}$ of the present sample is nearly zero ( = 2.3±4 mT), similarly to Sample II. However, very interestingly, in contrast to the invariant DW shape in Sample II, this sample exhibits clear DW tilting as shown by Fig. 4(b), even though both the samples have nearly zero $H_{\text{DMI}}$. Therefore, the DW tilting is not governed by $H_{\text{DMI}}$ for the present sample.

To explore the governing factor of the present DW tilting, the $v_{\text{DW}}$ variation with respect to $H_\perp$ is measured as shown by Fig. 4(c). The results clearly show that the present sample exhibits asymmetric $v_{\text{DW}}$ variation, in contrast again to the symmetric variation of Sample II. Such asymmetric $v_{\text{DW}}$ variation is possibly caused by the chirality dependent mechanisms such as the chiral damping [13, 14] and asymmetric DW width variation [15] even without the DMI. According to the previous discussion, such asymmetric $v_{\text{DW}}$ variation can cause the DW tilting via the difference of $v_{\text{DW}}$ between the upper and lower parts of the DW. Therefore, one can deduce that the DW tilting in the present sample is governed by the asymmetric $v_{\text{DW}}$ variation, which is possibly caused by the chirality-induced mechanisms [12-17].

Since the slope of the $v_{\text{DW}}$ variation near $H_\perp = 0$ is opposite to that of Sample I (Fig. 1(a)), the direction of the DW tilting of the present sample should be also opposite to that of Sample I. The present sample thus exhibits clockwise DW tilting, as confirmed by the experimental observation from Fig. 4(b). Therefore, it is general to conclude that the direction of the DW tilting is determined directly by the $v_{\text{DW}}$ asymmetry, whatever the origin of the asymmetry—either the DMI [9] or chirality-induced mechanisms [12-17]—is.

## DW tilting by both DMI and chirality-induced $v_{\text{DW}}$ asymmetry

Finally, as a general case, we examine the case that the sample has both a sizable DMI and chirality-induced $v_{\text{DW}}$ variation. Figure 5(a) plots $\varepsilon_{\text{ST}}$ with respect to $H_x$ for Sample IV (for sample details, see methods). The typical spin-orbit-torque-driven behavior again quantifies $H_{\text{DMI}}$ [23] as about $-110 \pm 5$ mT. This sample exhibits also the chirality-induced asymmetric $v_{DW}$ variation with respect to $H_\perp$ as shown by Fig. 5(b).

It is intriguing to see that the direction of the DW tilting is reversed depending on the magnitude of applied $H_y$. Figure 5(c) shows the observation results. The first two images were captured under application of a smaller $H_y$ (= 50 mT), whereas the last two images were captured under application of a larger $H_y$ (= 200 mT). It is clearly seen from the figure that the directions of the DW tilting are opposite between these two cases: counterclockwise DW tilting for the smaller $H_y$ in contrast to the clockwise DW tilting for the larger $H_y$. Figure 5(d) summarizes the DW tilting angle $\theta_{tilt}$ with respect to $H_y$. The plot shows that $\theta_{tilt}$ is reversed across a threshold magnetic field $H_y^{th}$ (green vertical line). Therefore, a counterclockwise DW tilting appears when $H_y < H_y^{th}$, otherwise a clockwise DW tilting appears.

The present peculiar results can be explained by considering the asymmetric $v_{DW}$ variation with respect to $H_\perp$. Recalling that $H_\perp$ at the upper part of the DW is opposite to the lower part of the circular DW arc as shown by Fig. 2(a), the initial difference $\Delta v_{DW}$ of the DW speed between the upper and lower parts of the DW basically follows the relation $\Delta v_{DW} = v_{DW}(H_\perp) - v_{DW}(-H_\perp)$. To visualize $\Delta v_{DW}$, $v_{DW}(-H_\perp)$ is plotted by the gray dashed line in Fig. 5(b) together with $v_{DW}(H_\perp)$ of the symbols with blue line. Then, $\Delta v_{DW}$ corresponds to the vertical difference between these two lines (purple arrow). It is again seen from the figure that the sign of $\Delta v_{DW}$ is reversed across a threshold magnetic field $H_\perp^{th}$ (green vertical line). Thus, for the case that $H_\perp$ is smaller than $H_\perp^{th}$, the DW rotates counterclockwise with $\Delta v_{DW} < 0$ and *vice versa*, in accordance to the observation from Fig. 5(d). Figure 5(e) shows the plot between $\theta_{tilt}$ and $\Delta(\ln v_{DW})$ for Sample IV. A clear correlation supports the validity of our model. Therefore, it is confirmed again that the direction of the DW tilting is truly determined directly by the asymmetry of the $v_{DW}$ variation.

**Discussion**

There remain several challenging issues towards full analytic description of the DW tilting angle $\theta_{tilt}$. First, we consider the effect of $H_\parallel$. Since the sign and magnitude of $H_\parallel$ at both the upper and lower parts of the DW are the same as seen by Fig. 2(a), the difference should be mainly attributed to the opposite sign of $H_\perp$. For the case near the Bloch-type configuration, by applying the first-order Taylor expansion with respect to $\psi$, the DW energy density $\sigma_{DW}$ becomes $\sigma_0 - \frac{1}{2}\pi\lambda M_S[(H_\perp + H_{DMI})^2 + 2H_S|H_\parallel|]/(H_S + |H_\parallel|)$ with the first leading terms of $H_\perp$ and $H_\parallel$, where the DW anisotropy field $H_S$ is defined as $4K_D/\pi M_S$. The difference $\Delta\sigma_{DW}$ is then given by $-\pi\lambda M_S H_\perp^+[H_{DMI}/(H_S + |H_\parallel|)]$, where $H_\perp^+$ denotes $H_\perp$ at the upper parts of the DW. Note that $\Delta\sigma_{DW}$ is directly proportional to $H_\perp^+$, but less sensitive to $H_\parallel$. Similarly, for the case near the Néel-type configuration with a large $H_{DMI}$, $\Delta\sigma_{DW} \cong -\frac{1}{2}\pi\lambda M_S H_\parallel[H_\parallel/(H_{DMI} + H_\perp)]$, which is negligible due to the small ratio of $H_\parallel/(H_{DMI} + H_\perp)$. Therefore, one can conclude that the direction of the DW tilting is mainly governed by the sign of $H_\perp$, whereas $H_\parallel$ might adjust the tilting angle slightly.

Second, though the mechanism discussed with $\Delta v_{DW}$ successfully explains the direction of the DW tilting as seen in Fig. 5(e), it is worthwhile to consider that there should be a restoring force to reach an equilibrium state with a finite $\theta_{tilt}$. It is clear that the DW tension induces the restoring force, since the DW tension energy increases as the DW tilts. The role of the tension-induced force can be roughly estimated as follows. The DW tension energy

$E_\text{tension}$ is given by $E_\text{tension} = \sigma_\text{DW} wt \sec\theta_\text{tilt}$ as a function of $\theta_\text{tilt}$ within the assumption of a straight DW, where $w$ and $t$ are the wire width and thickness, respectively. When the DW tilts from $\theta_\text{tilt}$ to $\theta_\text{tilt} + \delta\theta$, the upper and lower parts of the DW has the variation $\delta E^\pm_\text{tension}$ as given by $\frac{1}{2}\sigma_\text{DW} wt\delta\theta \sec\theta_\text{tilt} \tan\theta_\text{tilt}$, where the superscripts, $+$ and $-$, indicate the upper and lower parts of the DW, respectively. Similarly, the Zeeman energy has the variation $\delta E^\pm_\text{Zeeman}$ by $\pm\frac{1}{2}M_S H_z w^2 t\delta\theta \sec^2\theta_\text{tilt}$. Comparing these two energy variations, one can get the relation of the tension-induced effective magnetic field $H^\pm_z$ as $H^\pm_z = \pm\sigma_\text{DW} \sin\theta_\text{tilt}/M_S w$. For the case of a counterclockwise DW tilting (i.e. $\theta_\text{tilt} > 0$), a positive $H^+_z$ enhances the DW speed of the upper part while a negative $H^-_z$ reduces the DW speed of the lower part, resulting in clockwise rotation and *vice versa*. Therefore, the DW tension exerts a restoring force on the DW. The equilibrium $\theta_\text{tilt}$ is then determined by the steady-state condition i.e. $v_\text{DW}(H^+_\perp, H_z + H^+_z) = v_\text{DW}(H^-_\perp, H_z + H^-_z)$. Though an analytic solution of the present condition is not available due to the yet unknown nature of the chirality-induced $v_\text{DW}$ asymmetry, the empirical solution can be possibly provided by analyzing the two-dimensional map of $v_\text{DW}(H_\perp, H_z)$ measured as a function of $H_\perp$ and $H_z$.

In summary, we investigate here the DW tilting mechanism under application of in-plane magnetic field. For this study, four typical samples—Sample I (with DMI, but without chirality-induced asymmetry), Sample II (without both DMI and chirality-induced asymmetry), Sample III (without DMI, but with chirality-induced asymmetry), and Sample IV (with both DMI and chirality-induced asymmetry)—are examined. The experimental results clearly manifest that the DW tilting is triggered by the DW pinning at the structure edge and then, the direction of the DW tilting is governed by the DW speed asymmetry, whatever the origin of the asymmetry—either the DMI [9] or chirality-induced mechanisms [12-17]—is. The angle of the DW tilting is determined by the counterbalance between the DW speed asymmetry and the DW tension. A theory based on the DW speed asymmetry is provided.

**Method**

**Sample preparation.** The sample structures are 2.5-nm Pt/0.9-nm Co/2.5-nm Cu/1.5-nm Pt (Sample I), 4.0-nm Pt/0.3-nm Co/1.5-nm Pt (Sample II), 2.5-nm Pt/0.5-nm Co/1.5-nm Pt (Sample III), and 2.5-nm Pt/0.9-nm Co/2.5-nm Al/1.5-nm Pt (Sample IV), respectively. These samples are chosen as the representatives of typical major properties: Sample I (with DMI, but without chirality-induced asymmetry), Sample II (without both DMI and chirality-induced asymmetry), Sample III (without DMI, but with chirality-induced asymmetry), and Sample IV (with both DMI and chirality-induced asymmetry). These samples were deposited by use of dc magnetron sputtering on Si substrates with 100-nm-thick $SiO_2$ and 5-nm-thick Ta buffer layers. To enhance sharpness of the layer interfaces, each layer was deposited at a low deposition rate (~0.25 Å/s) under an Ar sputtering pressure (~2 mTorr).

**Observation of magnetic domain and DWs.** The magnetic domains are observed by use of a magneto-optical Kerr effect microscope which has $500\ \mu m \times 375\ \mu m$ field of view. The observed images are captured by a low light level CCD camera. To apply in-plane and out-of-plane magnetic field, 2-axis electromagnets are attached to the system. The maximum magnetic field strengths for each axis are 38 and 200 mT along the $z$ and $x$ axes, respectively. The two electromagnets are carefully aligned to avoid any possible cross-talk between two magnetic fields. At first the magnetization of the magnetic wire is saturated to the single down domain. Next, the up domain is nucleated by application of the local Oersted's field generated by injecting currents (up to 0.5 A) in to the gold wire, which is perpendicular to the magnetic wire. Then external magnetic field is applied to propagate the generated up domain. The image of the propagating domain is captured with a constant time interval to determine the speed of the DW.

**Acknowledgements**

This work was supported by grants from National Research Foundations of Korea (NRF) funded by the Ministry of Science, ICT and Future Planning of Korea (MSIP) (2015R1A2A1A05001698 and 2015M3D1A1070465). Y.-K.P. and B.-C.M. were supported by the National Research Council of Science & Technology (NST) (Grant No. CAP-16-01-KIST) by the Korea government (MSIP). D.-H.K. was supported from Overseas researcher under Postdoctoral Fellowship of Japan Society for the Promotion of Science (Grant No. P16314).

**Author contributions**

D.-Y.K. planned and designed the experiment and S.-B.C. supervised the study. D.-Y.K., J.-S.K., and Y.-S.N. performed the measurements. M.-H.P., Y.-K.P., H.-C.C. and B.-C.M. prepared the samples. S.-B.C., D.-Y.K., D.-H.K., and S.-G.J. performed the analysis and wrote the manuscript. All authors discussed the results and commented on the manuscript.

**Additional information**

Correspondence and request for materials should be addressed to S.-B.C.

**Competing financial interests**

The authors declare no competing financial interests.

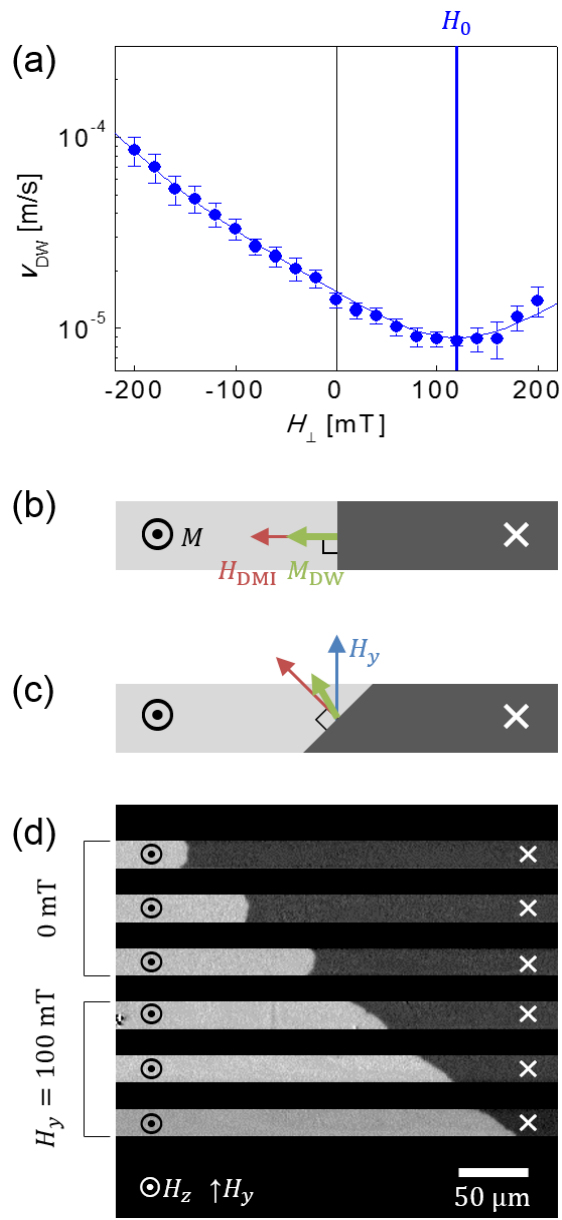

**Figure 1** (a) Plot of $v_{DW}$ as a function of $H_\perp$ for the Sample I. Blue vertical line indicates the compensate field $H_0$ for the DMI-induced effective field $H_{DMI}$. Schematic diagram of the DW for the magnetic wire with negatively large $H_{DMI}$ without (b) and with (c, 100 mT) application of $H_y$. Red, green, and blue arrows represent $H_{DMI}$, the magnetization inside the DW $M$, and applied $H_y$, respectively. (d) Image of the domains and DW for the micro-wire-patterned Sample I. First three and last three images show the propagating domain with and without application of $H_y$, respectively.

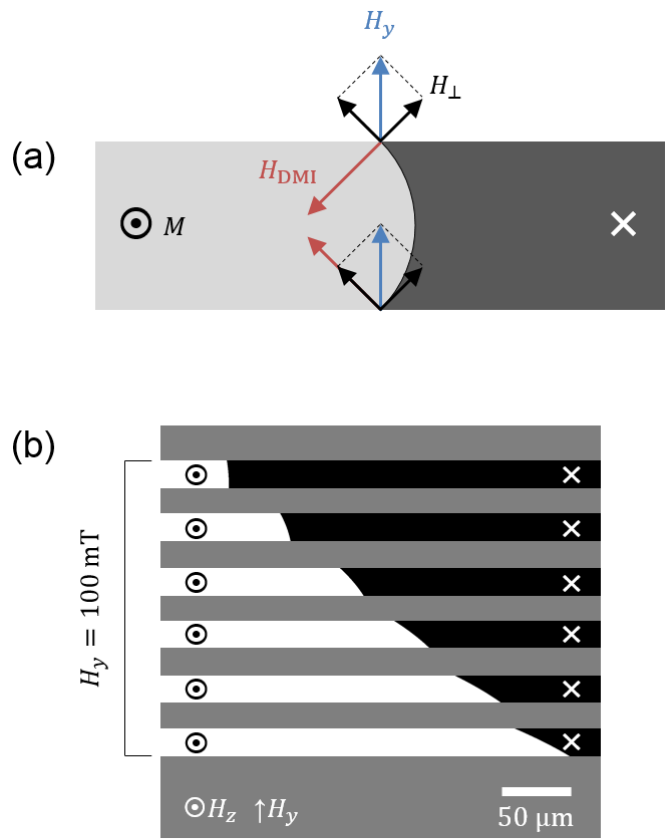

**Figure 2** (a) Schematic diagram of the arc-like-shaped DW due to the strong wire-edge pinning. Each red and blue arrows represent $H_{\mathrm{DMI}}$ and $H_y$. Black arrows indicate perpendicular and parallel component of $H_y$ ($H_\perp$ and $H_\parallel$). (b) Results of the numerical simulation.

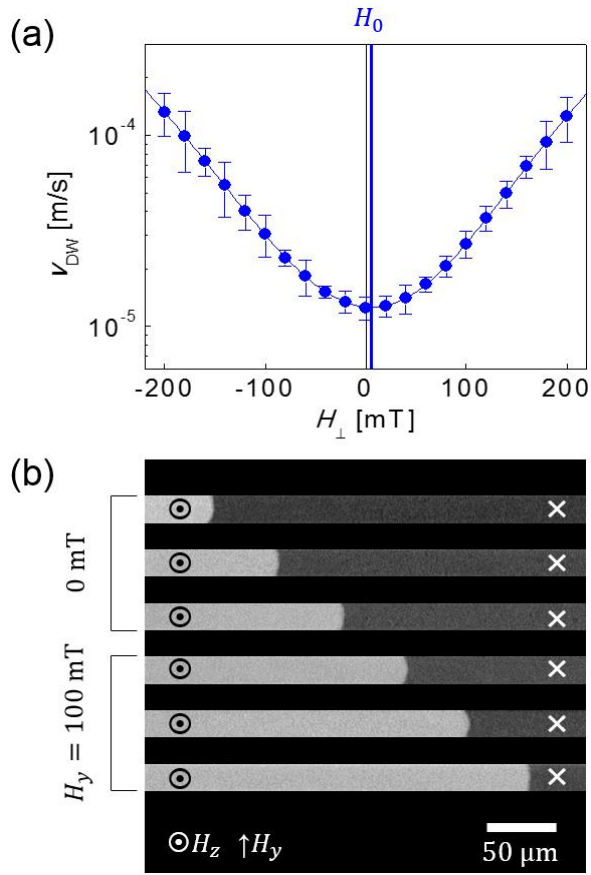

**Figure 3** (a) Plot of $v_{DW}$ as a function of $H_\perp$ for the Sample II. Blue vertical line indicates $H_0$. (b) Image of the domains and DW for the micro-wire-patterned Sample II. First three and last three images show the propagating domain without and with ($100 \text{ mT}$) application of $H_y$, respectively.

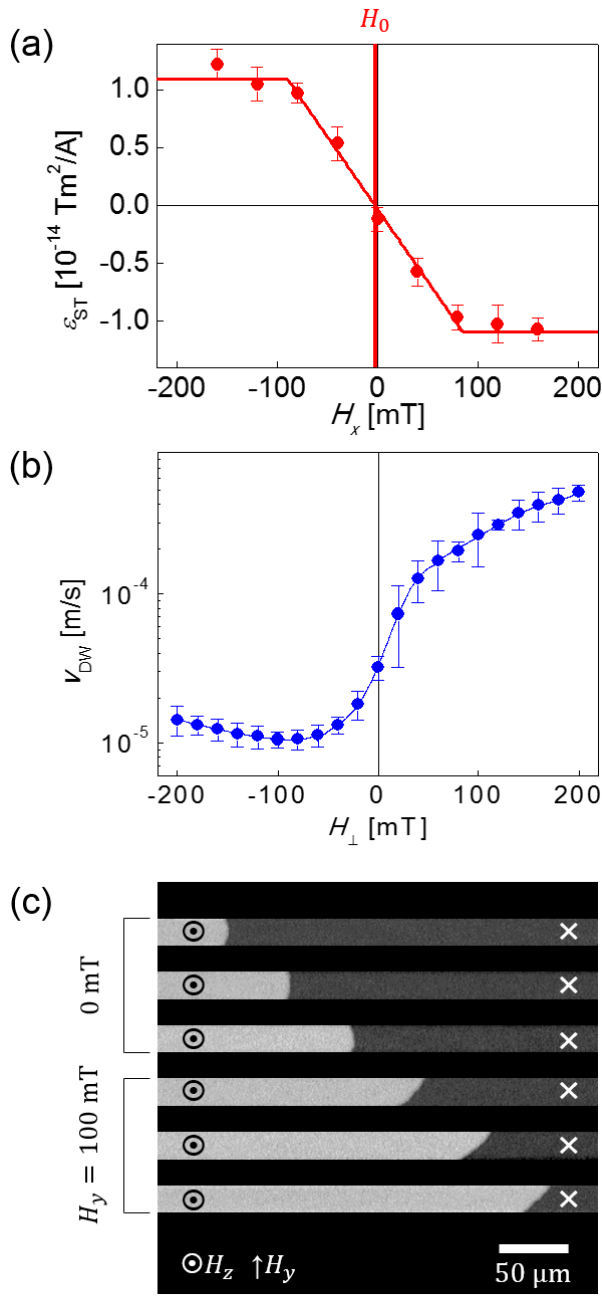

Figure 4 (a) Plot of spin-torque efficiency $\varepsilon_{ST}$ as a function of $H_x$ for the Sample III. Red vertical line indicates $H_0$. (b) (a) Plot of $v_{DW}$ as a function of $H_\perp$ for the Sample III. (c) Image of the domains and DW for the micro-wire-patterned Sample III. First three and last three images show the propagating domain without and with (100 mT) application of $H_y$, respectively.

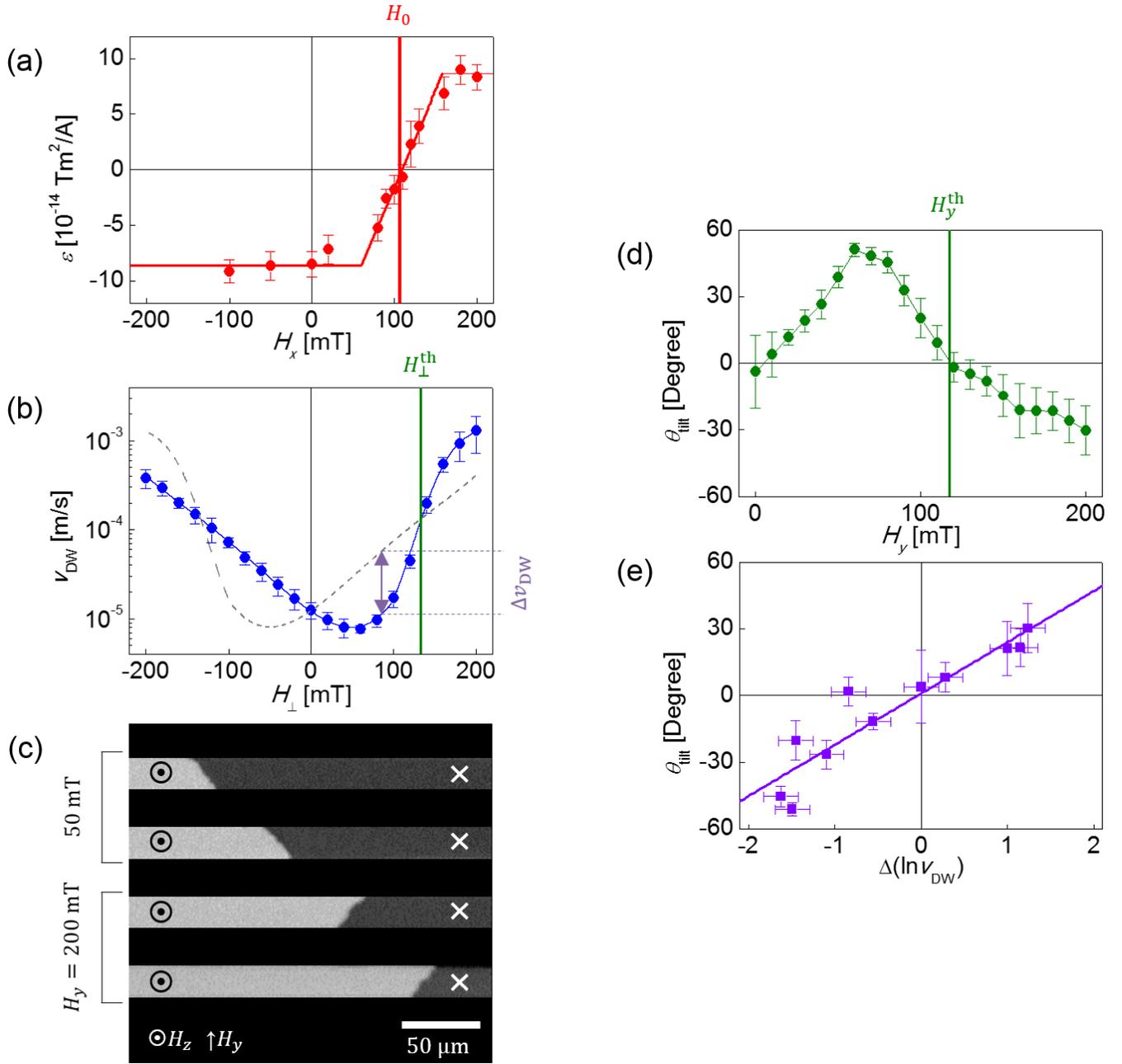

Figure 5 (a) Plot of spin-torque efficiency $\varepsilon_{ST}$ as a function of $H_x$ for the Sample IV. Red vertical line indicates $H_0$. (b) (a) Plot of $v_{DW}$ as a function of $H_\perp$ for the Sample IV. Gray dashed line represents $v_{DW}$ of the lower edge as a function of $H_\perp$. Green vertical line indicates $H_\perp^{th}$, where blue and dashed-gray curves intersect. (c) Image of the domains and DW for the micro-wire-patterned Sample IV. First two and last two images show the propagating domain under the application of $H_y = 50$ mT and $H_y = 200$ mT, respectively. (d) Plot of $\theta_{tilt}$ as a function of $H_y$. Green vertical line indicates $H_y^{th}$, where $\theta_{tilt} = 0$. (e) Plot of $\theta_{tilt}$ with respect to $\Delta(\ln v_{DW})$. The best linear fitting ($R^2 = 0.87$) is presented by the purple solid line.